\documentclass[review,number,sort&compress]{elsarticle}

\usepackage{graphicx}
\usepackage{epstopdf}
\usepackage{dcolumn}
\usepackage{amsmath}
\usepackage{mathptmx, courier, pifont}
\usepackage[scaled=0.92]{helvet}
\usepackage[T1]{fontenc}
\usepackage{textcomp}
\usepackage{color}
\usepackage{lineno,hyperref}
\begin{document}
%\draft
%\preprint{APS/123-QED}
%%%%
\begin{frontmatter}

\title{Experimental Investigation of the Transition Energy $\gamma_t$ in the Isochronous Mode of the HIRFL-CSRe}

%%%%
\cortext[my corresponding author]{Corresponding authors}
\author[A,B]{W.~W.~Ge}
\author[A]{Y.~J.~Yuan\corref{my corresponding author}}
\ead{yuanyj@impcas.ac.cn}
\author[A]{J.~C.~Yang}
\author[A]{R.~J.~Chen}
\author[A]{X.~L.~Yan}
\author[A]{H.~Du}
\author[A,B]{Z.~S.~Li}
\author[A]{J.~Yang}
\author[A]{D.~Y.~Yin}
\author[A]{L.~J.~Mao}
\author[A]{X.~N.~Li}
\author[A]{W.~H.~Zheng}
\author[A]{G.~D.~Shen}
\author[A,B]{B.~Wu}
\author[A,B]{S.~Ruan}
\author[A,B]{G.~Wang}
\author[A,B]{H.~Zhao}
\author[A]{M.~Wang}
\author[A]{M.~Z.~Sun}
\author[A]{Y.~M.~Xing}
\author[A,B]{P.~Zhang}
\author[A,B]{C.~Y.~Fu}
\author[A]{P.~Shuai}
\author[A]{X.~Xu}
\author[A]{Y.~H.~Zhang}
\author[A]{T.~Bao}
\author[A,C]{X.~C.~Chen}
\author[A,D]{W.~J.~Huang}
\author[A,B]{H.~F.~Li}
\author[A,B]{J.~H.~Liu}
\author[A,C]{Yu.~A.~Litvinov}
\author[A,C]{S.~Litvinov}
\author[E]{Q.~Zeng}
\author[A]{X.~Zhou}

\address[A] {Institute of Modern Physics, Chinese Academy of Sciences, Lanzhou 730000, China}
\address[B] {University of Chinese Academy of Sciences, Beijing 100049, China}
\address[C] {GSI Helmholtzzentrum f{\"u}r Schwerionenforschung, Planckstra{\ss }e 1, 64291 Darmstadt, Germany}
\address[D] {CSNSM, Univ Paris-Sud, CNRS/IN2P3, Universit\'{e} Paris-Saclay, 91405 Orsay, France.}
\address[E] {School of Nuclear Science and Engineering, East China University of Technology, Nanchang 330013, China}

\begin{abstract}
The Isochronous Mass Spectrometry (IMS) based on storage rings is a powerful technique for mass measurement of short-lived exotic nuclei.
The transition energy $\gamma_t$  of the storage ring is a vital parameter of the IMS technique.
It is difficult to measure the $\gamma_t$ and its relation to momentum spread or circulating length, especially to monitor the variation of $\gamma_t$ during experiments.
An experimental investigation on the $\gamma_t$ has been performed for the IMS experiment at the Cooler Storage Ring of the Heavy Ion Research Facility in Lanzhou (HIRFL-CSRe).
With the velocity measured by two time-of-flight (TOF) detectors, the $\gamma_t$ as a function of orbital length can be determined.
The influences of higher order magnetic field components on the $\gamma_t$ function were inferred for isochronous correction.
This paper introduces and investigates the influence of dipole magnetic fields, quadrupole magnetic fields and sextupole magnetic fields on the $\gamma_t$ function.
With the quadrupole magnets and sextupole magnets corrections, a mass resolution of 171332 (FWHM) and $\sigma(T)/T=1.34\times10^{-6}$ were reached, which shall be compared with 31319 (FWHM) and $\sigma(T)/T=7.35\times10^{-6}$ obtained without correction.

\end{abstract}

\begin{keyword}
%\texttt
{Isochronous Mass Spectrometry \sep transition energy \sep two-TOF detector system \sep nonlinear momentum compaction factor \sep first and second-order isochronous correction}
%\MSC[2010] 00-01\sep  99-00
%\pacs {21.10.Dr, 27.40.+z, 29.20.db}
\end{keyword}

\end{frontmatter}
\section{Introduction}
The accurate knowledge of nuclear masses plays an important role in nuclear structure physics and nuclear astrophysics \cite{Lunney2003,Blaum2006}.
Today, the challenge is to obtain precise masses of nuclei located far away from the valley of $\beta$-stability \cite{100yms}.
Since such nuclei are characterized by low production cross-section and short half-lives, highly efficient and fast mass measurement techniques are required.
The IMS based on storage rings is one of such techniques which are realized at in-flight radioactive ion beam facilities \cite{Litvinov2010,Bosch2013,ZhangPS2016}.
The IMS is performed today at three heavy-ion storage ring facilities in the world, namely the experimental storage ring ESR at GSI Helmholtz Center in Darmstadt \cite{Hausmann2001, Stadlmann2004, Sun2008, KnobelEPJ2016}, the experimental cooler-storage ring CSRe at the Institute of Modern Physics in Lanzhou \cite{Meng2009,TuNIMA2011,TuPL2011,ZhangPRL2012,YanAPL2013,ShuaiPLB2014,Xu2016,Zhang2017}, and the rare-ion storage ring R3 at the RIKEN Nishina Center in Tokyo \cite{Ozawa2012,Yamaguchi2013a,Yamaguchi2013b,Wakasugi2015,Yamaguchi2015a,Yamaguchi2015b}.

The principle of the IMS is expressed by Eq. (1), which connects the revolution time ($T$) of stored ions to their mass-to-charge ratio ($m/q$) and velocity ($v$) \cite{WollnikNIMA1987,HausmannNIMA2000,Franzke2008}:
\begin{equation}
   \frac{\Delta T}{T} = \frac{1}{{\gamma_{t}}^{2}}\frac{\Delta (m/q)}{m/q} + \Biggl(\frac{{\gamma}^{2}}{{\gamma_t}^{2}}-1\Biggr)\frac{\Delta v}{v},
   \label{eqims}
\end{equation}
where $\gamma$ is the relativistic Lorentz factor and $\gamma_t$ is the transition energy of the storage ring.
The transition energy is defined as:
$\gamma_t = 1/{\sqrt{\alpha_p}}$, where $\alpha_p$ is the momentum compaction factor.

For definite magnetic rigidity and transition energy settings of a storage ring, only ions with certain $m/q$ can satisfy the ``isochronous condition'' $\gamma=\gamma_t$.
To differentiate, these ions can be called as ``isochronous ions''.
In isochronous condition, the second term on the right hand side of Eq. (1) can be reduced to a negligible value and the $T$ only depends on $m/q$.
For storage rings, the $\gamma_t$ usually is not a constant value within the full momentum acceptance, so the isochronous condition is not always completely fulfilled and the velocity spread still contributes to the revolution time deviation $\Delta T$.

In order to increase the mass resolving power, there are three methods to reduce $\Delta T/T$  \cite{ChenRST2015}.
First, one can tune the slope of the dependence of $\gamma_t$ on $\Delta v/v$ to reduce the phase slip factor $\eta$ ($\eta=1/\gamma^{2}-1/\gamma_t^{2}$) \cite{LitvinovNIMA2013}.
Second, one can reduce the $\Delta v/v $  by using so-called $B\rho$-tagging method  \cite{KnobelPLB2016,Geissel2005}.
The mass resolving power can be significantly increased though at a cost of a dramatically reduced transmission.
Third, one can correct the revolution time by measuring their velocities \cite{DolinskiiNIMA2007,XingPST2015,XuCPC2015}.
In principle, this method has a great advantage to improve the mass resolving power of IMS without losing valuable statistics.
In all the approaches above, the dependency of $\gamma_t$ on momentum deviation plays an important role.
Therefore, it is critical to investigate the $\gamma_t$ for the improvement of the mass resolution.

Researchers have made great efforts to improve the isochronicity and chromaticity of the IMS technique.
In 2000, Hausmann et al. have investigated the properties of the isochronous mode of the ESR.
They found that the experimental variation of $\gamma_t$($\Delta B\rho$) is due to higher order variations of the dispersion and the mean gradient of $\gamma_t$($\Delta B\rho$) can be controlled by the sextupoles \cite{HausmannNIMA2000}.
S.Litvinov et al. found that the nonlinear lattice without any higher order correction has a very strong effect on the revolution frequency spread, which results in a low mass resolving power.
According to their simulations, sextupoles are required to correct chromaticity and reduce the influence of higher order field imperfections and fringe fields.
Additionally, sextupoles, octupoles and decapoles are necessary to correct the pure momentum dependence of isochronicity correction in higher orders \cite{LitvinovNIMA2013}.
X. Xu et al. presented an isochronous mass measurement method using a two-TOF detector system.
The simulation results show that the new scheme can significantly improve mass resolving power with the additional velocity information of stored ions.
This improvement is especially important for nuclei with $\gamma$ value far away from the $\gamma_t$ \cite{XuCPC2015}.
X.L. Yan et al. used the Schottky Mass Spectrometry (SMS) measured frequencies to investigate the shape of the $\alpha_p$ ($\alpha_p=1/\gamma_t^{2}$) of the ESR.
They found that the knowledge of $\alpha_p$ shape is important for accurate calibration of Schottky frequency spectra and directly affects the systematic errors of the measured masses \cite{Yan X LPS2015}.

Although the IMS technique has been studied intensively, precise measurement of the relation between transition energy $\gamma_t$ (or momentum compaction factor $\alpha_p$) and the momentum deviation $\Delta p/p$ (or closed-orbit length $L$) is still a big challenge.
Recently, R.J. Chen et al. introduced a new method to accurately measure the $\gamma_t$ function as $L$\cite{chenruijiu2016}.
Compared with the previous method of using electron cooler, the new method with two TOF detectors has an obvious advantage.
The $\gamma_t$ curve can be measured online during mass measurement experiments.

In this paper, the new method mentioned above was used to measure the $\gamma_t$ function in the IMS of the HIRFL-CSRe.
The effects of dipole magnetic fields, quadrupole magnetic fields and sextupole magnetic fields on the $\gamma_t$ function were investigated.
By applying the corrections of quadrupole magnets and sextupole magnets, the mass resolving power of the target nuclei reached 171332 (FWHM) and $\sigma(T)/T=1.34\times10^{-6}$.

\section{The New Isochronous Mode with Two TOF Detectors of the HIRFL-CSRe }
The CSRe, an experimental Cooler Storage Ring of the Heavy Ion Research Facility in Lanzhou (HIRFL-CSR) complex, is a high precision spectrometer which is operated with different detectors for the experiments on atomic physics with highly charged ions and nuclear spectroscopy measurements of the radioactive secondary nuclei  \cite{Xia2008}.
As shown in Figure 1, the CSRe is a two-fold symmetric ring with a total circumference of 128.8 meters.
As one of the three operation modes of the CSRe, the isochronous mode is operated for the IMS experiments of very short-lived exotic nuclei.
In early IMS experiments, only one TOF detector was used to measure the revolution time of stored ions.
The employed transition energy settings of the CSRe were 1.395, 1.302 and 1.280.
The $\beta$-function and dispersion function of the $\gamma_t=1.395$ are shown in Figure 2 (a).
The relative revolution time difference was only $10^{-6}$ in the setting of $\gamma_t=1.395$ \cite{GX2014}.

The two TOF detectors were installed in one straight section of the CSRe.
The momentum of stored ions can be properly measured with the additional velocity information, and thus the mass resolving power can be significantly improved.
%With the two TOF detectors, the isochronous optics first needs to fulfil the injection condition of the CSRe.
%Then the dispersion at the detector positions needs to be as small as possible to increase the momentum acceptance.
For the two TOF scheme, a new isochronous optics has to be designed.
The optics should match the injection line of the CSRe to warrant a high transmission.
Meanwhile, the dispersion should be small at the positions where the two detectors are installed, which is a prerequisite for a reasonably large momentum acceptance.
A new optics setting with $\gamma_t=1.360$ was accommodated to meet these two conditions.
The $\beta$-function and dispersion function of the new optics are shown in Figure 2 (b).

Considering the aperture of the vacuum pipe of the CSRe, the new optics allow us to store beams with a momentum acceptance of $\pm0.3\%$ and a transverse emittance of $30$ $\pi$ $mm\cdot mrad$ in each plane.
Figure 3 shows the theoretical $\gamma_t$ and  $\gamma$ as functions of the momentum deviation of stored ions in the linear lattice of the CSRe.
As illustrated in Figure 3, the isochronous condition is not strictly fulfilled in the linear lattice calculations when the momentum deviation from the central momentum increases.
Obviously, higher order isochronous corrections are needed to improve the resolving power of the measured revolution times.
\begin{figure}\centering
    \includegraphics[angle=0,width=8.5 cm,height=6 cm]{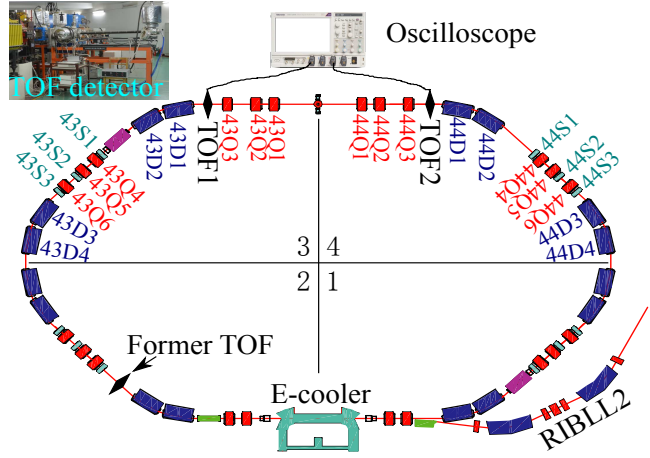}
	\caption{
(Colour online) The layout of the CSRe.
The blue, red and cyan rectangles indicate the dipole magnets, quadrupole magnets and sextupole magnets, respectively.
	\label{DistributionOfC}}
\end{figure}
\begin{figure}\centering
	\includegraphics[angle=0,width=8.5 cm,height=6 cm]{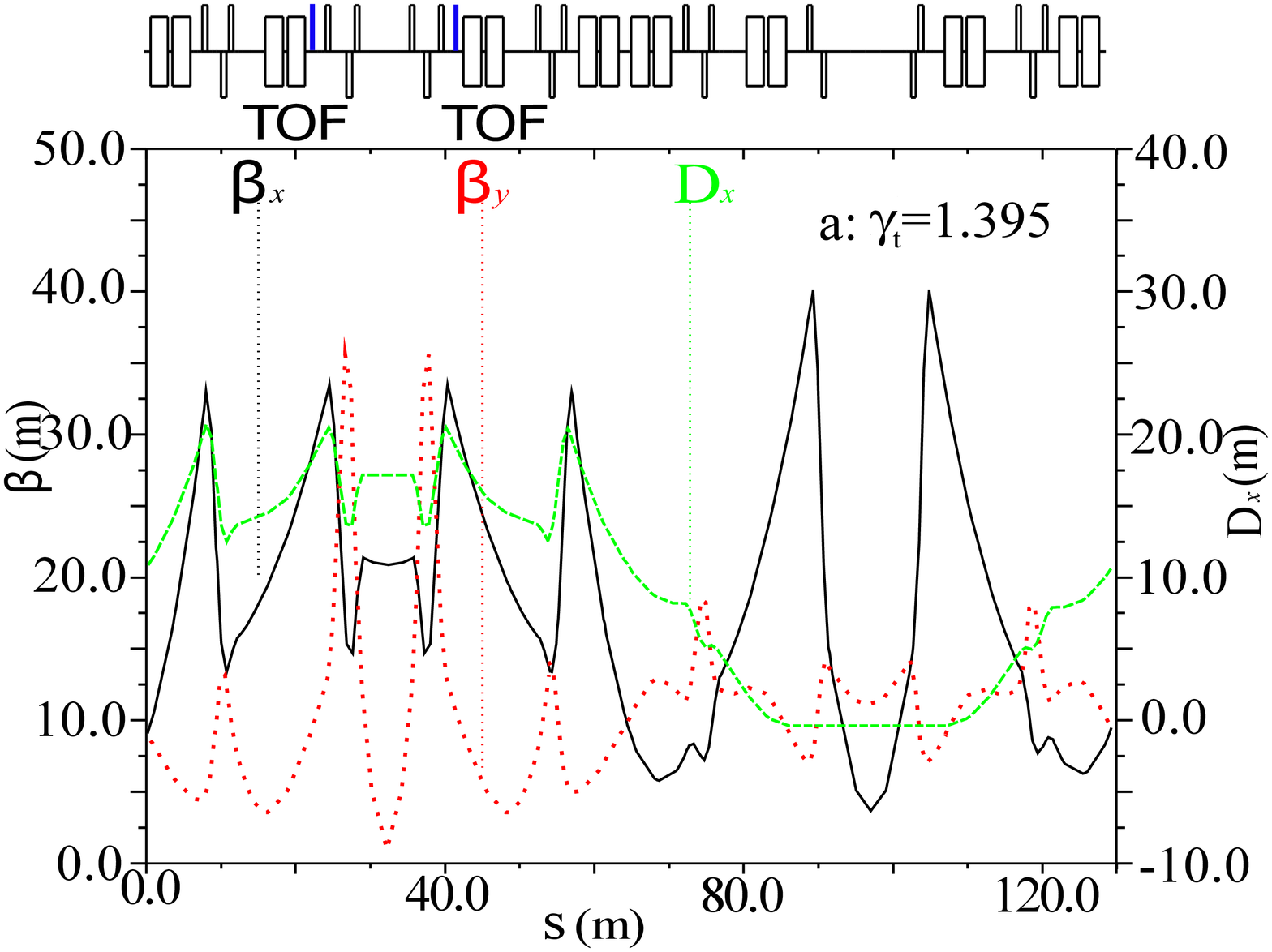}
    \includegraphics[angle=0,width=8.5 cm,height=6 cm]{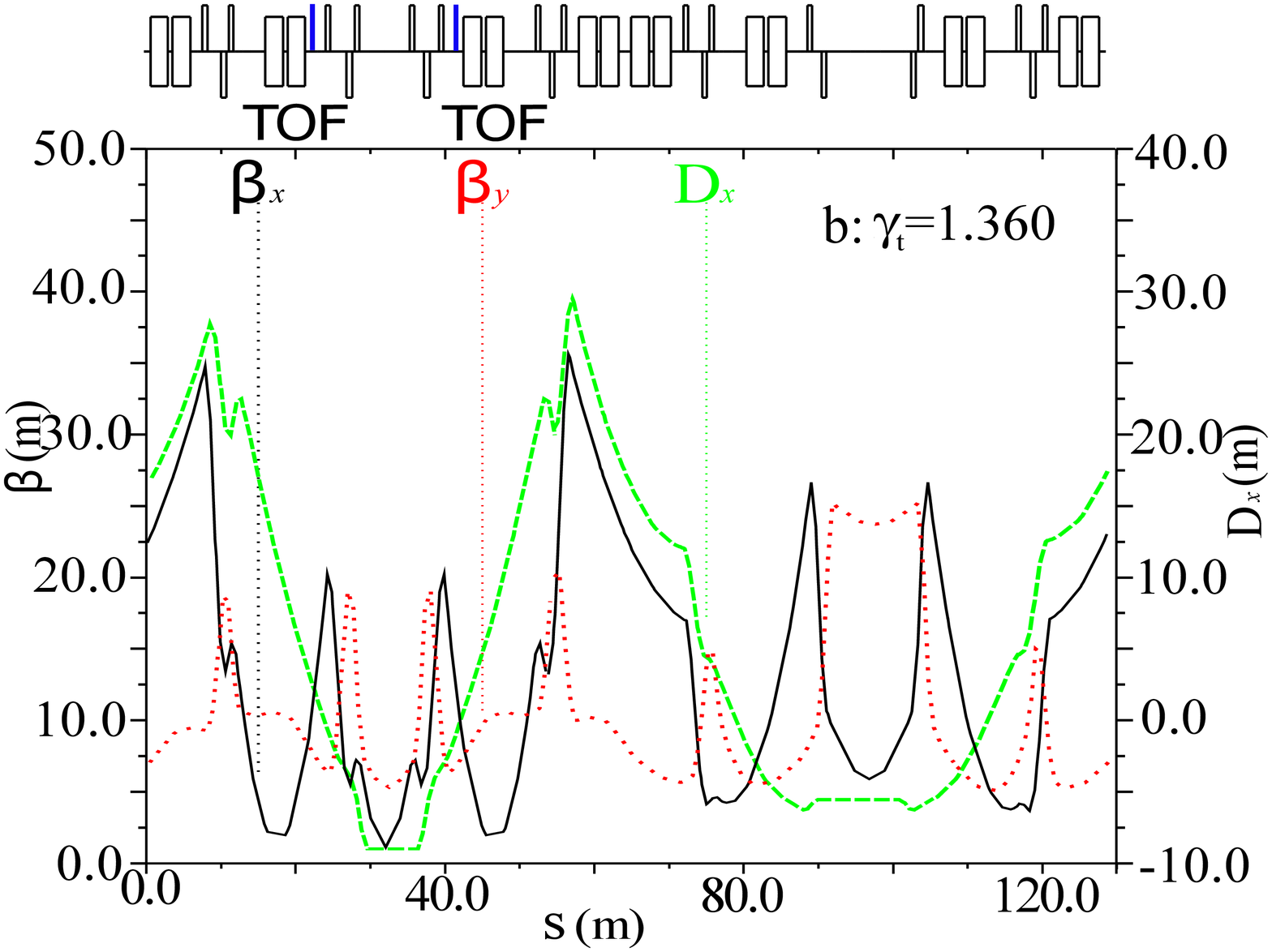}
	\caption{
(Colour online) The $\beta$-functions and dispersion function as a function of the orbit length.
The thick black line, red dotted line and green dashed line represent the $\beta_x$, $\beta_y$ and $D_x$ function, respectively.
%The initial position of the figure is the center of the electron cooler.
The positions where the two TOF detectors are installed are indicated with blue rectangles.
%Panels (a) and (b) show the data for the $\gamma_t=1.395$ and $1.360$, respectively.
	\label{DistributionOfC}}
\end{figure}
\begin{figure}\centering
	\includegraphics[angle=0,width=8.5 cm,height=6 cm]{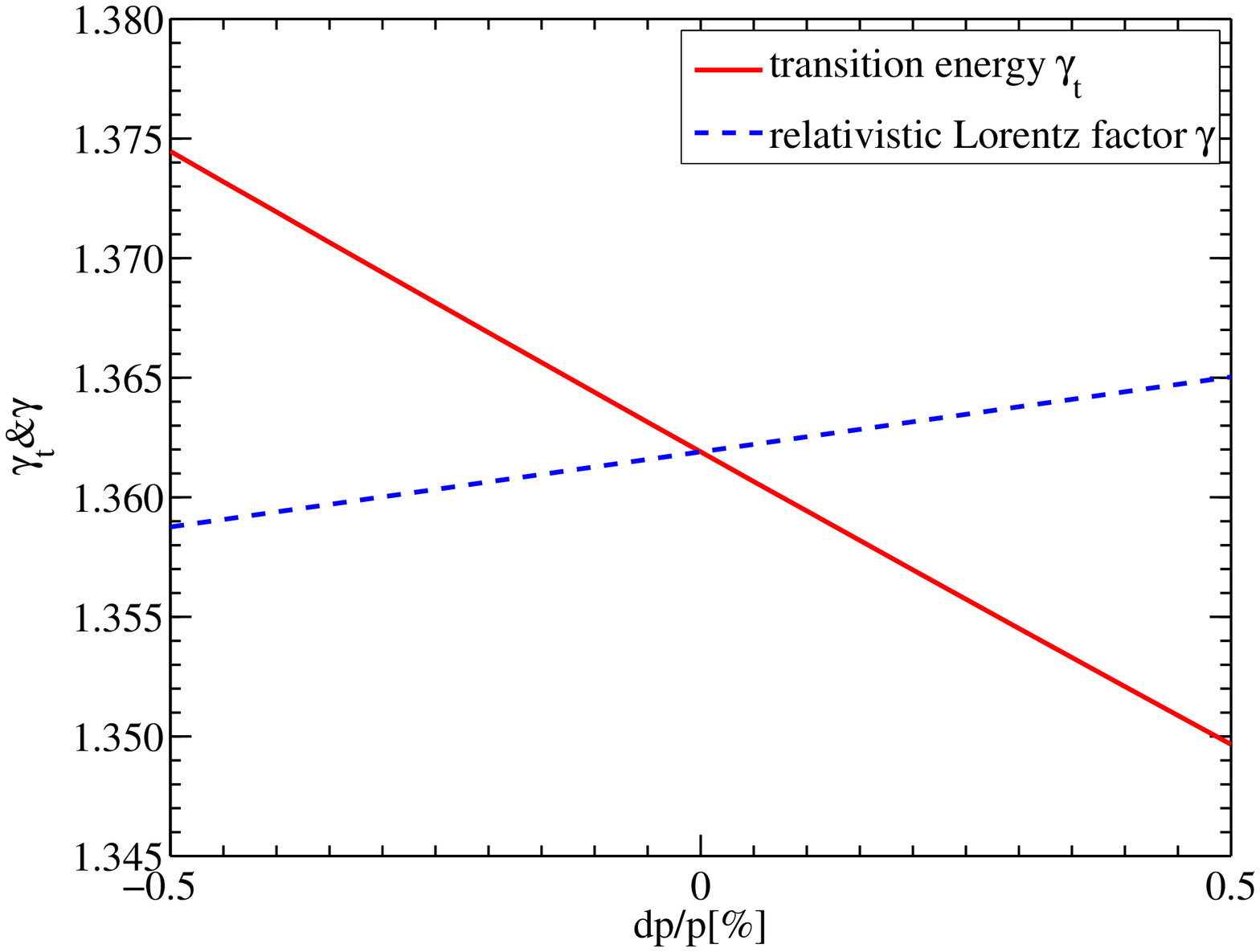}
	\caption{
	(Colour online) The theoretical $\gamma_t$ and $\gamma$ as a function of the momentum deviation in the linear lattice of the CSRe.
	\label{DistributionOfC}}
\end{figure}

\section{Nonlinear Momentum Compaction Factor and Phase Slip Factor}
%%In order to describe the non-linear profile of the $\gamma_t$ function, we deduced basic formulas in this section.
The basic formulas of nonlinear profile of the $\gamma_t$ function are introduced in this section.
The $\gamma_t$ is deduced from the momentum compaction factor $\alpha_p$ of the storage ring as $\gamma_t= \sqrt{1/\alpha_p}$.
The various orders of momentum compaction factor $\alpha_i$ give the relative spread in closed-orbit length $L_0$ for an off-momentum particle \cite{Delahaye1985,Shan1992}:
\begin{equation}
\frac{\Delta L}{L_0} =\int_{0}^s{\sqrt{(1+\frac{\Delta x}{\rho(s)})^{2}+(\Delta x^{'})^{2}}ds}= \alpha_p\delta=\sum_{i=0}^n \alpha_i\delta^{i+1}
\label{eqims}
\end{equation}
where $\Delta x=\sum_{i=0}^n D_i(s)\delta^{i+1}$, $\Delta x^{'}=\sum_{i=0}^n D_i^{'}(s)\delta^{i+1}$, $s$ denotes the coordinate along the reference orbit $L_0$ in the ring, $\rho(s)$ is the bending radius of the reference orbit, and $\delta=(p-p_0)/p_0$ is the momentum spread.
$D(s)$ and $D'(s)$ are the dispersion function and its derivative.
The index $i$ indicates the order of the Taylor expansion.

The the various orders of $\alpha_i$ can be expressed as follows:
\begin{eqnarray}
\alpha_0&=&\frac{1}{L_0}\int_{0}^s\frac{D_0(s)}{\rho(s)}\,ds\nonumber\\
\alpha_1&=&\frac{1}{L_0}\int_{0}^s\frac{D_1(s)}{\rho(s)}\,ds+\frac{1}{2L_0}\int_{0}^s\frac{D_0'(s)^2}{\rho(s)}\,ds\\
\alpha_2&=&\frac{1}{L_0}\int_{0}^s\frac{D_2(s)}{\rho(s)}\,ds+\frac{1}{L_0}\int_{0}^s\frac{D_0'(s)D_1'(s)}{\rho(s)}+\frac{1}{2L_0}\int_{0}^s\frac{D_0(s)D_0'(s)^2}{\rho(s)}\,ds\nonumber
\label{eqims}
\end{eqnarray}

For the same species of ion, the first term in Eq. (1) vanishes and the relative revolution time difference can be expressed as a function of various orders of $\eta_i$:
\begin{equation}
\frac{\Delta T}{T_0} =(\frac{L}{v}-\frac{L_0}{v_0})\frac{1}{T_0}= \eta\delta=\sum_{i=0}^n \eta_i\delta^{i+1}
\label{eqims}
\end{equation}
By combining Eq. (2) and (4), we obtain:
\begin{eqnarray}
\eta_0&=&\alpha_0-\frac{1}{\gamma_0^2}\nonumber\\
\eta_1&=&\alpha_1-\frac{\eta_0}{2\gamma_0^4}+\frac{3(\gamma_0^2-1)}{2\gamma_0^4}\\
\eta_2&=&\alpha_2-\frac{\eta_1}{2\gamma_0^4}+\frac{3(\gamma_0^2-1)\eta_0}{2\gamma_0^4}+\frac{4\gamma_0^4-9\gamma_0^2+5}{2\gamma_0^6}\nonumber
\label{eqims}
\end{eqnarray}

As illustrated in Eq. (5), the nonlinear phase slip factor $\eta$ depends mainly on the $\alpha_i$ and the $\gamma_0$.
The $\gamma_0$ depends on the $m/q$ of isochronous ions.
The $\alpha_i$ are determined by the storage ring optics.
As shown in Eq. (3), the $\alpha_i$ are produced by the higher order components of the magnetic field, and the central value is determined by $D_0$ and $\rho(s)$.
Then $\eta_i= 0$ as so-called ``isochronous condition'' could be achieved through the corrections of quadrupole magnets, sextupole magnets, octupole magnets, and so on.

\section{The Influence of Magnetic Fields on the Transition Energy $\gamma_t$}
An experiment verifying the influence of magnetic fields on the $\gamma_t$ was performed with the IMS in the CSRe.
In the experiment, a primary ${}^{40}Ar^{14+}$  beam at 7.0 $MeV/u$ provided by the HIRFL was accumulated, cooled, and further accelerated to 389.2 $MeV/u$ in the CSRm.
The projectile fragments were produced via primary beam bombarding a 15 $mm$ beryllium target which was installed at the entrance of the RIBLL2 fragment separator.
The secondary beams were separated and purified with the RIBLL2 and then injected into the CSRe.
As isochronous ions ${}^{38}S^{16+}$ were selected, the RIBLL2-CSRe system was set to achieve the optimal transmission for the secondary beams, which corresponded to $B\rho$ = 6.8065 $Tm$ and $\gamma_t=1.360$.

In the IMS experiment, the circulating ions lose energy each time when they penetrate carbon foils of the two TOF detectors \cite{chenruijiu2016,MeiNIMA2010}.
The change of their velocity and revolution time can be measured during a storage time, which is about 200 microseconds.
Using these information, one can obtain the $\gamma_t$ experimental expression as follows:
\begin{equation}
\gamma_t=\frac{1}{\sqrt{\frac{1}{\gamma^{2}}+2\frac{\gamma}{\gamma+1}\frac{E_k}{\Delta E_k}\frac{A_2}{T}}}
\label{eqims}
\end{equation}
where $\Delta E_k$ is the average energy loss and $E_k$ the kinetic energy.
$A_2$ equals to the revolution time change per two revolution turns of stored ions in the ring as measured by the detectors.

The investigation on the $\gamma_t$ functions are described in the following sections.
In the first step, the relation of the $\gamma_t$ curve to the orbit length was measured with the original setting with $\gamma_t=1.360$, which will be served as a reference.
Then the dipole magnetic fields, quadrupole magnetic fields and sextupole magnetic fields were varied, the $\gamma_t$ curve was measured to verify the influence of these changes on the $\gamma_t$.
Due to the variation or instabilities of magnetic fields, the measured $\gamma_t$ represents the averaged over the measurement period.
Figures (4-8) show the $\gamma_t$ curve as a function of the orbital length.
QC and SC mean quadrupole magnet correction and sextupole magnet correction, respectively.
$B$, $k$ and $\lambda$ are the magnetic field strength of dipole magnets, quadrupole magnets and sextupole magnets, respectively.
The index $0$ indicates the initial isochronous setting with $\gamma_t=1.360$.

%The range of the optimal region for the experiment is $\delta \gamma_t/\gamma_t<0.35\%$.
In the perfect setting for the IMS experiments, the $\gamma_t$ curve should behave exactly the same as the gamma curve of the target nucleus.
However, in practice such an ideal condition cannot be achieved.
In the IMS experiments we try to control the fluctuation of the $\gamma_t$ within the momentum acceptance to obtain high mass resolution.
If the fluctuation is too large within the full momentum acceptance, we use experimental data only located within a restricted region where the fluctuation is acceptable for a reasonable mass resolution.
In the investigation, we define the working region with $\delta (\gamma_t)/\gamma_t<0.35\%$, then the spread of the revolution time $\sigma(T)/T<1\times10^{-5}$ can be achieved.

\subsection{The Influence of Dipole Magnets on the Transition Energy $\gamma_t$}

%When all dipole magnetic fields were changed by $\pm0.2\%$, two adjustments of the CSRe were performed.
All the dipole magnets of the CSRe are connected in series and served with one power supply.
By tuning that power supply, the dipole magnetic fields were increased by $0.2\%$ and the $\gamma_t$  curve was measured, and then the fields were decreased by $0.2\%$  for another measurement.
As illustrated in Figure 4, increasing or decreasing the dipole magnetic fields setting caused the $\gamma_t$ curve to move left or right in the horizontal direction, respectively.
In the same orbit, the $\gamma_t$ changed by 0.01, for example, in the orbit length of 128.85 meters, the $\delta\gamma_t/\gamma_{t0}=(\gamma_t-\gamma_{t0})/\gamma_{t0}= 0.73\%$.
In addition, the change of dipole magnetic fields setting resulted in a reduction in the optimal region for the experiment.
%When the dipoles increased by 0.002 times, the region was changed from ``128.83-128.94'' to ``128.87-128.91'', compared with which were reduced 0.002 times, the region became ``128.83-128.91''.
As shown in Table 1, when the dipoles setting was changed, the optimal region for the experiment is reduced on the right side.
The maximum reduction of the optimal region for the experiment is $27.27\%$.

As revealed in Eq. (3), when the dipole magnetic fields increase, then the $\rho$ decreases and the $\alpha_p$ increases, so the $\gamma_t$ curve moves downwards.
Similarly, when the dipole magnetic fields decrease, the $\gamma_t$ curve moves upwards.

The results show that outside of the optimal region for the experiment, the experimental data are different from the theoretical values.
The latter is due to the distribution of the fringing fields, misalignments of the magnets and the magnetic measurement errors, which are also related to the instability of the magnetic field.
This feature can also be seen in Figures (5-8).
Therefore, a more accurate actual magnetic field distribution and a more stable power supplies are needed to provide accurate data for theoretical calculations.

\begin{figure}\centering
	\includegraphics[angle=0,width=8.5 cm,height=6 cm]{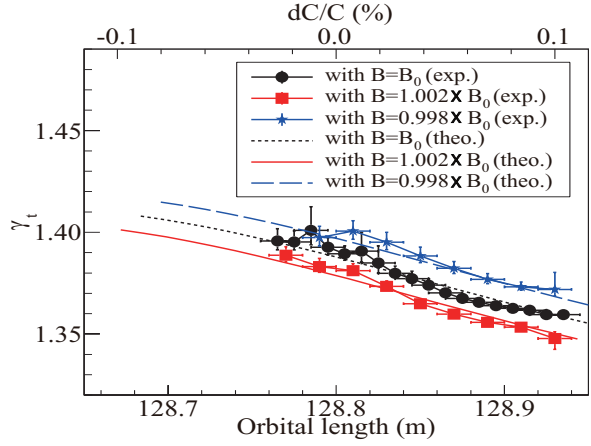}
	\caption{
	(Colour online) The transition energy $\gamma_t $ as a function of circulating length due to variations of dipole magnetic fields setting.
	\label{DistributionOfC}}
\end{figure}

\begin{table}[h]
\centering
\begin{tabular}{|l|c|c|}\hline
dipole magnetic fields setting&the optimal region[m]\\\hline
$B=B_0$ &128.83-128.94\\\hline
$B=1.002\times B_0$  &128.83-128.91\\\hline
$B=0.998\times B_0$  &128.83-128.91\\\hline
\end{tabular}
\caption{The influence of dipole magnetic fields setting on the optimal region for the experiment.}
\label{tab:Margin_settings}
\end{table}

\subsection{The Influence of Quadrupole Magnets on the Transition Energy $\gamma_t$}

%When the magnetic fields of one quadrupole family (43Q6 and 44Q6) in the largest dispersion position were changed by $\pm1\%$, the CSRe was adjusted in two performances.
A pair of quadrupole magnets, 43Q6 and 44Q6 as shown in Fig.1, which are located at the positions where the dispersion is the largest, are used to study the influence of the quadruple magnets on the CSRe performance.
The fields of these two magnets were changed by $\pm1\%$ and the corresponding performances were measured.
As shown in Figure 5, increasing or decreasing the chosen quadrupole magnetic fields made the $\gamma_t$ curve to move downwards or upwards in the vertical direction, respectively.
In the same orbit, the $\gamma_t$ is changed by 0.035, for example, the $\delta\gamma_t/\gamma_{t0}= 2.54\%$ at the orbit length of 128.85 meters.
%In addition, the change of quadrupoles setting did not caused change in the optimal region for the experiment.
%The region was changed from ``128.83-128.94'' to ``128.83-128.95'' when the selected quadrupoles were increased by 0.01 times, compared with reduced 0.01 times, and the region became ``128.75-128.87''.
As shown in Table 2, the chosen quadrupole magnetic fields setting changed by $\pm1\%$, the optimal region for the experiment was moved, but the size was not changed.

As revealed in Eq. (3), when the chosen quadrupole magnetic fields increase, then the ability of changing dispersion increases.
In other words, $D_0$ and $D_0^{'}$ are reduced, the $\alpha_0$ and $\alpha_1$ are reduced, and the $\gamma_t$ curve shifts upwards.
Similarly, when the chosen quadrupole magnetic fields decrease, the $\gamma_t$ curve shifts downwards.

\begin{figure}\centering
	\includegraphics[angle=0,width=9 cm,height=6 cm]{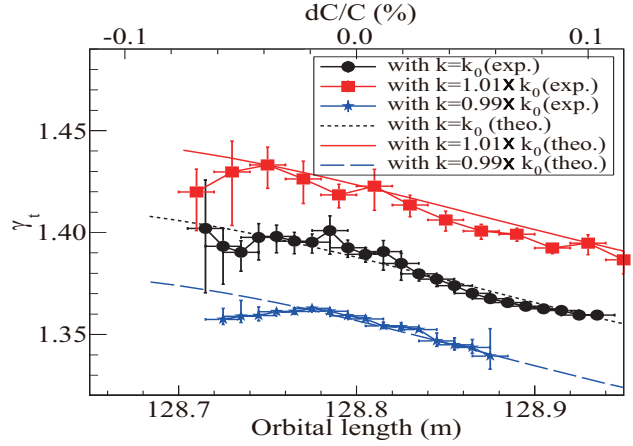}
	\caption{
	(Colour online) The transition energy $\gamma_t $ as a function of the circulating length due to variations of the chosen quadrupole magnetic fields setting.
	\label{DistributionOfC}}
\end{figure}

\begin{table}[h]
\centering
\begin{tabular}{|l|c|c|}\hline
43Q6,44Q6 &the optimal region[m]\\\hline  % \hline 在此行下面画一横线
$k=k_0$ &128.83-128.94\\\hline        % \\ 表示重新开始一行     % & 表示列的分隔线
$k=1.01\times k_0$  &128.83-128.95\\\hline
$k=0.99\times k_0$  &128.75-128.87\\\hline
\end{tabular}
\caption{The influence of the chosen quadrupole magnetic fields setting on the optimal region for the experiment.}
\label{tab:Margin_settings}
\end{table}

\subsection{The Influence of Sextupole Magnets on the Transition Energy $\gamma_t$}

%When the magnetic fields of one sextupole family (43S1 and 44S1) were modified by 0 and 2 times its original value, the CSRe was adjusted in two properties.
A pair of sextupole magnets, 43S1 and 44S1 as shown in Fig.1, are used to study the influence of the sextupole magnets on the CSRe performance.
The fields of these two magnets were changed by 0 and 2 times its original value and the corresponding performances were measured.
As shown in Figure 6, increasing and reducing the chosen sextupole magnetic fields rotated the $\gamma_t$ curve anticlockwise or clockwise, respectively.
In addition, the change of the chosen sextupole magnetic fields setting resulted in the reduction of the optimal region for the experiment.
%The region was changed from ``128.70-128.85'' to ``128.70-128.81'' when the selected sextupoles were increased by 2 times, while reduced to 0, and the region became ``128.70-128.77''.
As shown in Table 3, when the chosen sextupole magnetic fields setting increased from 0, the right side of the optimal region for the experiment increased first and then decreased, and the maximum reduction of the optimal region for the experiment is $46.67\%$.

According to Eq. (3), when the chosen sextupole magnetic fields increase, then $D_1$ and $D_1^{'}$ reduce, the $\alpha_1$ and $\alpha_2$ reduce, so the $\gamma_t$ rotates anticlockwise.
Similarly, when the magnetic fields of chosen sextupole magnetic fields decrease, the $\gamma_t$ curve rotates clockwise.

\begin{figure}\centering
	\includegraphics[angle=0,width=8.5 cm,height=6 cm]{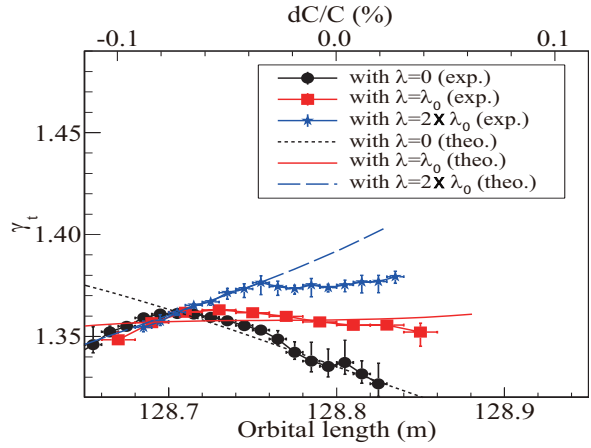}
	\caption{
	(Colour online) The transition energy $\gamma_t $ as a function of the circulating length due to variations of the chosen sextupole magnetic fields setting.
	\label{DistributionOfC}}
\end{figure}

\begin{table}[h]
\centering
\begin{tabular}{|l|c|c|}\hline
43S1, 44S1 &the optimal region[m]\\\hline  % \hline 在此行下面画一横线
$\lambda=0$  &128.70-128.77\\\hline
$\lambda=\lambda_0$ &128.70-128.85\\\hline        % \\ 表示重新开始一行     % & 表示列的分隔线
$\lambda=2\times \lambda_0$  &128.70-128.81\\\hline
\end{tabular}
\caption{The influence of the chosen sextupole magnetic fields setting on the optimal region for the experiment.}
\label{tab:Margin_settings}
\end{table}

\section{Isochronous Correction}
For the isochronous mass measurement experiments, one important task is to make the isochronous condition $\gamma_t=\gamma$ fulfilled for the target nuclei over the momentum acceptance as much as possible, in order to decrease the isochronous deviations.
The isochronous deviations come from chromatic and geometric aberrations \cite{Karl1985}.
The second-order chromatic aberrations and geometric aberrations are caused by the imperfections of the dipole magnetic fields and sextupole magnetic fields, while pure quadrupole magnetic fields only produce chromatic aberrations.
The geometric aberrations can be corrected with a repetitive symmetric structure.
The chromatic aberrations can be corrected with high-order magnets.
In the IMS experiments of the CSRe, the various orders of the isochronicity deviations could be corrected by decreasing $\eta_i$ to zero with quadrupole magnet and sextupole magnet, etc.
The experimental results were described as follow.

\begin{table}[h]
\centering
\begin{tabular}{|l|c|c|}\hline
magnets name &43Q6,44Q6&43Q4,44Q4\\\hline  % \hline 在此行下面画一横线
initial setting($1/m^2$) &-0.3463864&-0.3537617\\\hline        % \\ 表示重新开始一行     % & 表示列的分隔线
QC1($1/m^2$) &-0.3438588&-0.3537617\\\hline
QC2($1/m^2$) &-0.3463864&-0.3500206\\\hline
\end{tabular}
\caption{The quadrupole magnetic fields setting for the first-order isochronous correction.}
\label{tab:Margin_settings}
\end{table}

\begin{figure}\centering
	\includegraphics[angle=0,width=8.5 cm,height=6 cm]{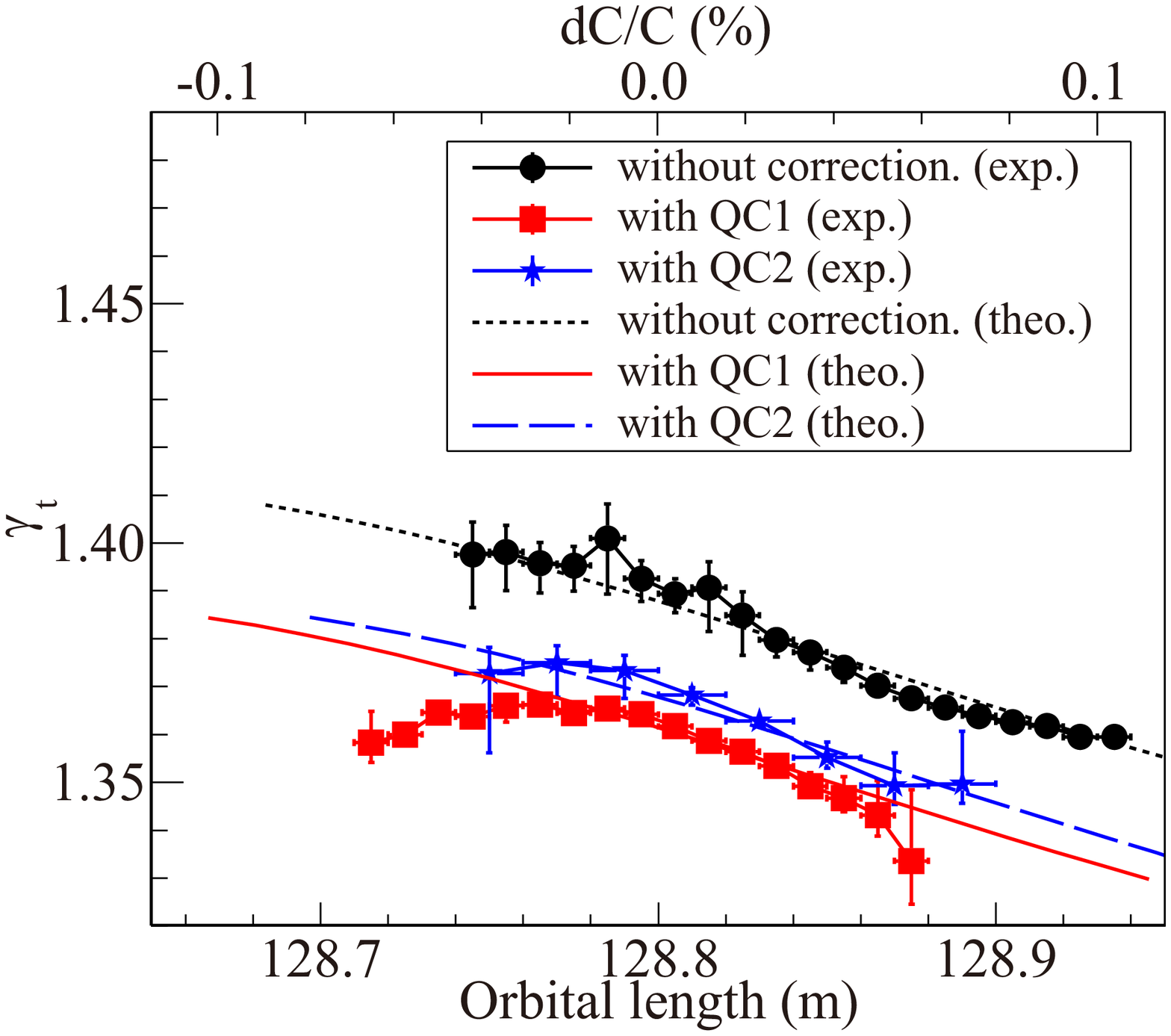}
	\caption{
	(Colour online) The transition energy $\gamma_t $ as a function of the circulating length due to isochronous corrections of quadrupole magnets.
	\label{DistributionOfC}}
\end{figure}

\begin{figure}\centering
	\includegraphics[angle=0,width=8.5 cm,height=6 cm]{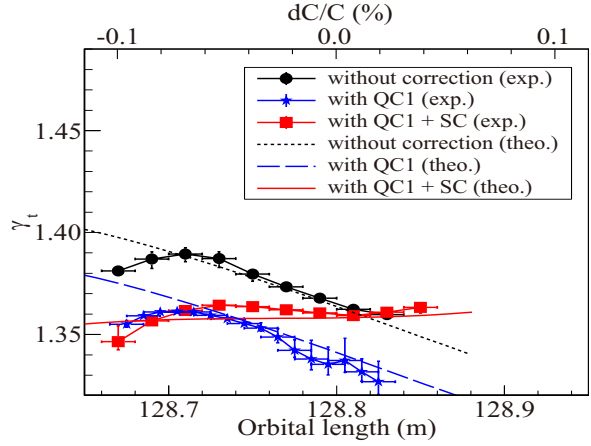}
	\caption{
	(Colour online) The transition energy $\gamma_t$ as a function of the circulating length due to isochronous corrections of quadrupole magnets and sextupole magnets.
	\label{DistributionOfC}}
\end{figure}

\subsection{ First-order Isochronous Correction}
Two quadrupole magnet families at different dispersions were applied to correct the first-order isochronous condition.
Table 4 shows the setting of the chosen quadrupole magnetic fields.
QC1 means only one family quadrupole magnet of 43Q6 and 44Q6, while QC2 only 43Q4 and 44Q4.
As displayed in Figure 7, although quadrupole magnet corrections at different dispersions have different effects, $\eta_1=0$ was approximately fulfilled.
The best option is to select the quadrupole magnet at the largest dispersion position.
The quadrupole magnet corrections marked QC1 can move the $\gamma_t$ curve downwards and make the $\gamma_t$ more closely to the designed valve of 1.360.
In the optimal region for the experiment ``128.75-128.87 $m$'', the mass resolving power of the target nuclei reached 36351 (FWHM) and $\sigma(T)/T=6.32\times10^{-6}$, compared with 31319 (FWHM) and $\sigma(T)/T=7.35\times10^{-6}$ of without correction.

\begin{figure}\centering
	\includegraphics[angle=0,width=8.5 cm]{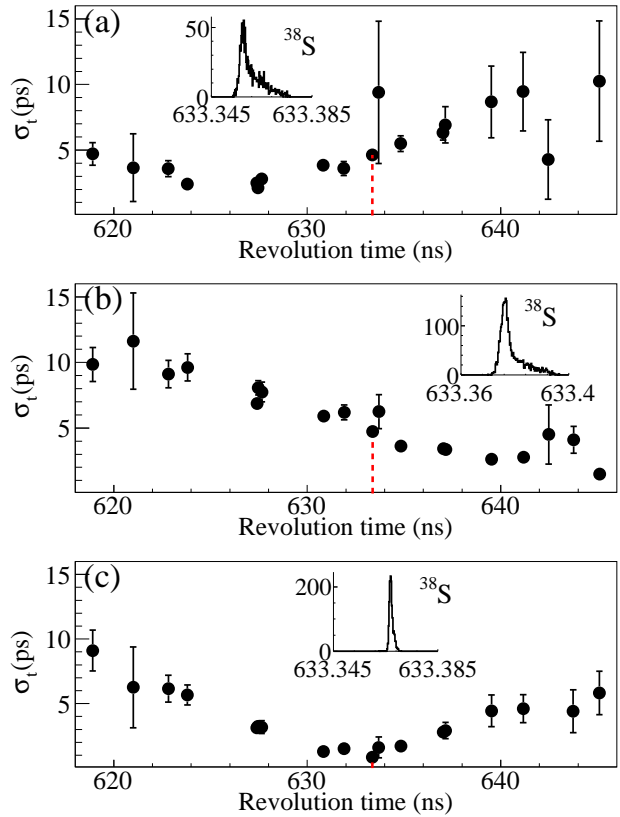}
	\caption{
	(Colour online) Standard deviations of revolution time distributions of different nuclides as a function of their mean revolution times.
Panels (a), (b) and (c) show the data for settings without corrections, with QC1 and with QC1+SC, respectively.
The inserts in figure are the revolution time distributions of ${}^{38}S^{16+}$ ions.
The red dashed line mean the corresponding isochronous setting where $\gamma_t=\gamma$ in the first order approximation.
	\label{DistributionOfC}}
\end{figure}

\subsection{ Second-order Isochronous Correction}
The second-order isochronous correction was performed by applying one sextupole magnet family ($43S1$ and $44S1$) at a maximum dispersion position.
The chosen sextupole magnetic fields were set on 0.003513625 $1/m^{3}$.
As illustrated in Figure 8, with quadrupole magnet corrections and sextupole magnet corrections marked by QC1+SC, $\eta_2=0$ was reached.
The $\gamma_t$ curve approximately became a flat line in the orbital length region of ``128.70-128.85 $m$'', and the mass resolving power of the target nuclei reached 171332 (FWHM) and $\sigma(T)/T=1.34\times10^{-6}$.

The characterization of the isochronous correction could also be reflected by the standard deviations of revolution time distributions of all nuclides versus their mean revolution times.
The standard deviation of the revolution time of ${}^{38}S^{16+}$ with the settings without corrections, QC1 and QC1+SC are presented in Figure 9.
The best isochronous ions are found at 625.522 $ns$, 639.469 $ns$ and 633.345 $ns$, respectively.
With quadrupole magnet and sextupole magnet corrections, the standard deviation of the revolution time decreases, and hence the mass resolving power improves.

\section{Conclusion}
The influence of different magnets on the $\gamma_t$ has been investigated employing the IMS in the isochronous mode of the CSRe.
The $\gamma_t$ curve was horizontally, vertically shifted as well as rotated, by varying dipole magnetic fields, quadrupole magnetic fields and sextupole magnetic fields, respectively.
Correspondingly, these adjustments changed the optimal region for the experiment by different ratios.
These results could be seen as a reference to control the $\gamma_t$ curve.
Combining the online monitor of the $\gamma_t$, we could tune the $\gamma_t$ curve more effectively during the IMS experiment.

The isochronous condition of the CSRe was greatly satisfied with the quadrupole magnets corrections and sextupole magnets corrections, and the mass resolving power of the target nuclei reached 171332 (FWHM) and $\sigma(T)/T=1.34\times10^{-6}$.
In order to reduce the impact of nonlinear fields on the isochronicity, the sextupole magnet, octupole magnet and perhaps higher order magnet are needed.
In the future, the application of the high-order isochronicity and chromaticity corrections with sextupole magnet and octupole magnet will be considered and applied for the IMS in the CSRe.
In addition, it is observed that high-order isochronicity and chromaticity corrections as well as two-TOF detectors system will also be implanted at the collector ring CR at the Facility for Antiproton and Ion Research FAIR in Germany  \cite{LitvinovNIMA2013} and the Spectrometer Ring at the High Intensity heavy-ion Accelerator Facility HIAF in China \cite{YJC2013}.

\section*{Acknowledgments}
This work is supported in part by National Key S$\&$T Program of China (Grant No. 2016YFA0400504),
the NSFC (Grant Nos.11605252, U1232208, U1432125, 11205205, 11035007, 11605248),
the Helmholtz-CAS Joint Research Group HCJRG-108,
the External Cooperation Program of the CAS (GJHZ1305),
the 973 Program of China (No. 2013CB834401),
and by the European Research Council (ERC) under the European Union’s Horizon 2020 research and innovation programme (grant agreement No. 682841 ``ASTRUm'').

\end{document}